# MOM: Matrix Operations in MLIR

Towards Compiler Support for Linear Algebra Computations in MLIR

Lorenzo Chelini
Huawei Technologies
Switzerland
l.chelini@icloud.com

Henrik Barthels
RelationalAI, Inc.
Germany
henrik.barthels@relational.ai

Paolo Bientinesi
Umea University
Sweden
paolo.bientinesi@umu.se

Marcin Copik
ETH Zurich
Switzerland
marcin.copik@inf.ethz.ch

Tobias Grosser
University of Edinburgh
UK
tobias.grosser@ed.ac.uk

Daniele G. Spampinato
Huawei Technologies
Switzerland
daniele.giuseppe.spampinato@huawei.com

**Abstract**

Modern research in code generators for dense linear algebra computations has shown the ability to produce optimized code with a performance which compares and often exceeds the one of state-of-the-art implementations by domain experts. However, the underlying infrastructure is often developed in isolation making the interconnection of logically combinable systems complicated if not impossible. In this paper, we propose to leverage MLIR as a unifying compiler infrastructure for the optimization of dense linear algebra operations. We propose a new MLIR dialect for expressing linear algebraic computations including matrix properties to enable high-level algorithmic transformations. The integration of this new dialect in MLIR enables end-to-end compilation of matrix computations via conversion to existing lower-level dialects already provided by the framework.

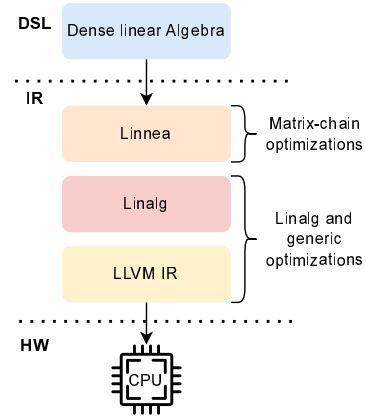

**Figure 1.** The MOM compiler for dense linear algebra.

## 1 Introduction

A significant part of processor time is spent on mathematical algorithms used in simulations, machine learning, communication, signal processing, computer vision, and other domains. Of course, the mathematics used in these domains may differ widely. Still, the actual computations often fall into the realm of linear algebra, meaning sequences of computations on matrices and vectors. Research in the area of linear algebraic domain-specific languages (DSLs) has demonstrated that expert-level optimizations can be carried out automatically when taking the mathematical semantics of the computation into account (e.g., [2, 7, 9]). Over time, however, the code generation infrastructure supporting such research has grown into a fragmented software ecosystem, where different components that could logically cooperate towards a shared performance goal are in practice unable to interoperate with one another. Thanks to the new MLIR [8] compiler infrastructure is now possible to de-fragmentize our software ecosystem by enabling domain-specific compilation directly in a general-purpose compiler. However, after careful evaluation of the available dialects in MLIR, none was able to capture linear-algebra semantics, including Linalg [10]; therefore, we developed our dialect.

This paper proposes an implementation of a subset of the state-of-the-art linear algebra code generator Linnea [2] in MLIR. Specifically, our contributions are:

- An IR representation to model linear algebra operations, types and properties.
- A lowering from such representation to LLVM IR.

## 2 The Linnea Dialect

We introduce a new linear algebraic dialect in MLIR called Linnea after the homonymous code generator [2]. Linnea provides MLIR attributes, types, operations, and transformations required to make dense computations as first-class citizens in the compiler IR. In Fig. 1 we show an overview of the Matrix Operations in MLIR (MOM [1]) compiler framework. Linnea bridges dense linear algebra computations expressed in our Python DSL and the Linalg dialect, as shown in Fig. 1. From Linalg, we lower to LLVM IR and then machine code. Linnea comes with custom attributes, types and operations described next.

---

[1]https://github.com/Algebraic-Programming/MLIR-Linnea

Lorenzo Chelini, Henrik Barthels, Paolo Bientinesi, Marcin Copik, Tobias Grosser, and Daniele G. Spampinato
Lorenzo Chelini, Henrik Barthels, Paolo Bientinesi, Marcin Copik, Tobias Grosser, and Daniele G. Spampinato

*Matrix Attribute.* Inspired by the Sparse Tensor Dialect [3], we introduce Matrix attributes to encode non-conflicting, compile-time properties of a matrix. For example, the property that a matrix is lower triangular is expressed using the following attribute:

```
#linnea.property<['lowerTri']>
```

During bufferization [2], the underlying algebraic object is materialized using a straightforward memref where attributes dictate how the memory is filled (i.e., set to zeros the elements below the diagonal for an upper triangular matrix). In the future, we aim to depart from memref and use more appropriate containers.

*Types: Matrix, Term and Identity.* Linnea provides three built-in types: Matrix, Term, and Identity. Matrix represents a 2-dimensional tensor type which accepts an attribute that describes a set of non-conflicting properties, a static dimension, and an MLIR built-in type that describes each stored element's type (e.g. f32). For example, the following is the type declaration of a $5 \times 5$ lower-triangular matrix:

```
!linnea.matrix#linnea.property<['lowerTri', [5, 5], f32]>
```

A term type represents the result of Linnea equationOp, which in turn models a mathematical equation, composable by the available operations (see next paragraph). A term is is a placeholder type, meaning that it will be replaced by a concrete one (i.e., MatrixType). In fact, only after optimization and simplification the result of an equationOp is known and thus the placeholder types can be swapped with a concrete one. Finally, an Identity type represents the identity matrix – a square matrix where all the elements on the diagonal are 1. The Identity type enables the expression of mathematical identities such as $A \cdot I = A$ and $I \cdot I = I$.

*Operations.* Table 1 shows the operations exposed by the Linnea dialect. Init and Fill initialize and fill an algebraic object, respectively. Mul and Add are variadic operations that take as input an arbitrary numbers of MLIR Values of Linnea types [6]. The Equation represents a "bag" of Linnea operations that together represent a given mathematical expression. We build a symbolic representation of the program by walking each equation starting from the yield operation. An equation is simplified and rematerialized to new low-level Linnea IR (not shown). For example, variadic multiplications are rewritten as binary multiplications with an optimal parenthesization.

*Transformations.* Currently, we limit ourselves to three transformations: matrix-chain reordering, identity simplification and properties propagation. Matrix chain reordering is an algorithmic improvement that minimizes the number of scalar multiplications when multiplying a chain of matrices. We implement the algorithm described in [5] and generalize

---

[2]bufferization materializes memref (i.e., malloc) from tensors.

| Op name | Description |
|---|---|
| Init | Materialize the algebraic object using a memref |
| Fill | Initialize the algebraic object according to its properties |
| Equation | Represents a linear algebra equation |
| Add | Variadic addition of different algebraic objects |
| Yield | Return the result of an Equation |
| Mul | Variadic multiplication of different algebraic objects |
| Transpose | Transpose the algebraic object |

**Table 1.** Operations exposed by the Linnea dialect.

```
1  n = 5
2  m = 5
3  Matrix A(n, m) <LowerTriangular>
4  Matrix B(n, m) <LowerTrinagular>
5  Matrix C(n, m) <>
6  C = A * B
7  print(C)
```

**Listing 1.** MOM specification for a triangular matrix multiplication.

it to account for matrix properties as in [1, 2]. Identity simplification consists of exploiting the identity matrix to simplify the computation. For example, $A \cdot I \rightarrow A$. Finally, Linnea's IR makes it possible to annotate matrices with properties. But, it is also essential to understand the properties of intermediate results as the computation unfold, thus we introduce a symbolic engine and encode a set of inference rules such as $lowerTriangular(A) \rightarrow upperTriangular(A^T)$. We use the symbolic engine to replace a TermType with a concrete type.

## 3 MOM Compiler Usage

MOM provides users with a convenient Python DSL language to express their computation. It does not require any knowledge about the internal intermediate representation or the different compiler passes involved in lowering a Python specification to binary. For example, Listing 1 shows how a user can specify a multiplication between two lower triangular matrices.

Behind the scene, MOM lowers the Python specification to the Linnea dialect (see Listing 2). Line 3 and 4 map to the initialization for the A and B matrix. Line 6 maps to the linnea.equation operation.

*Lowering to Linalg.* The conversion to Linalg is relatively straightforward; each operation in the Linnea dialect maps one-to-one with available operations in Linalg. For example, Listing 3 shows the Linalg operations materialized after the conversion of Listing 2. Matrix types are converted to two-dimensional tensors with the same element type and sizes. The attribute that specifies the type of the matrix is preserved.

```
// Initialize matrix A. Default initialize to 1 (%fc = 1).
%A = linnea.init [5, 5] :
  !linnea.matrix#linnea.property<['lowerTri', [5, 5], f32]>
%Af = linnea.fill(%fc, %A) :
  f32, !linnea.matrix#linnea.property<['lowerTri', [5, 5], f32]>
// Initialize matrix B.
%B = linnea.init [5, 5] :
  !linnea.matrix#linnea.property<['lowerTri', [5, 5], f32]>
%Bf = linnea.fill(%fc, %B) :
  f32, !linnea.matrix#linnea.property<['lowerTri', [5, 5], f32]>
// Multiply the two and print the result.
%0 = linnea.equation {
  %1 = linnea.mul %Af, %Bf :
    !linnea.matrix#linnea.property<['lowerTri', [5, 5], f32]>
    !linnea.matrix#linnea.property<['lowerTri', [5, 5], f32]>
    -> !linnea.term
  linnea.yield %1 : !linnea.term
}
linnea.print %0 : !linnea.term
```

**Listing 2.** A Linnea IR representation for a multiplication between two lower triangular matrices.

```
// Initialize matrix A. Default initialize to 1 (%fc = 1).
%A = linalg.init-tensor [5, 5]
  : tensor<5x5xf32, #linnea.property<['lowerTri']>>
%Af = linalg.fill ins(%fc : f32)
  outs(%A : tensor<5x5xf32, #linnea.property<['lowerTri']>>)
// Initialize matrix B.
%B = linalg.init-tensor [5, 5]
  : tensor<5x5xf32, #linnea.property<['lowerTri']>>
%Bf = linalg.fill ins(%fc : f32)
  outs(%B : tensor<5x5xf32, #linnea.property<['lowerTri']>>)
%C = linalg.init-tensor [5, 5]
  : tensor<5x5xf32, #linnea.property<['lowerTri']>>
linalg.matmul ins(%Af : tensor<5x5xf32, #linnea.property<['lowerTri']>>
                  %Bf : tensor<5x5xf32, #linnea.property<['lowerTri']>>)
  outs(%C : tensor<5x5xf32, #linnea.property<['lowerTri']>>)
linnea.print %C : tensor<5x5xf32, #linnea.property<['lowerTri']>>
```

**Listing 3.** A Lianlg IR representation for Listing 2.

## 4 Results

The experiments have been conducted on an Intel Core i7-10750H. All results were obtained considering the minimal execution time of five independent runs for single-precision (`f32`) operands as in [4]. For a chain of 4 matrices $800 \times 1100 \times 900 \times 1200 \times 100$ with initial parenthesization $(((A_1 \times A_2) \times A_3) \times A_4)$ we obtain a $1.24X$ speedup by re-parenthesizing as $(A_1 \times (A_2 \times (A_3 \times A_4)))$.

## 5 Conclusion

We presented MOM, an end-to-end compiler for dense linear algebra operations based on MLIR. Due to the missing algebraic semantics in Linalg, we proposed a new linear algebraic dialect for capturing essential matrix properties and enabling the expression of domain-expert optimization strategies. For example, it is effortless to detect matrix-chain multiplications thanks to Linnea and the introduction of variadic multiplications and matrix types. Future work includes the integration of more properties and algebraic transformations, as well as the investigation of suitable extensions of the existing infrastructure to support them (e.g., the multiplication of two triangular matrices could map to a non-rectangular iteration space currently not supported by Linalg).